\documentclass[aps,prl,showpacs,twocolumn,groupedaddress]{revtex4}
\usepackage[latin1]{inputenc}
\usepackage{graphicx}
\usepackage{color}

\begin{document}


\title{Disorder and relaxation mode in the lattice 
dynamics of \\PbMg$_{1/3}$Nb$_{2/3}$O$_3$ relaxor ferroelectric}



\author{S.N. Gvasaliya}
\altaffiliation{On leave from Ioffe Physical Technical Institute, 
26 Politekhnicheskaya, 194021, St. Petersburg, Russia}
\affiliation{Laboratory for Neutron Scattering ETHZ \& Paul-Scherrer
Institut 
CH-5232 Villigen PSI Switzerland}

\author{S.G. Lushnikov}
\affiliation{Ioffe Physical Technical Institute, 26 Politekhnicheskaya,
194021, 
St. Petersburg, Russia}

\author{B. Roessli}
\affiliation{Laboratory for Neutron Scattering ETHZ \& Paul-Scherrer
Institut 
CH-5232 Villigen PSI, Switzerland}

\date{\today}

\begin{abstract}
The low-energy part of vibration spectrum in PbMg$_{1/3}$Nb$_{2/3}$O$_3$  
relaxor ferroelectric was studied by inelastic neutron scattering. 
We observed the coexistence of a resolution-limited central peak with   
strong quasielastic scattering. The line-width of the quasielastic component 
follows a $\Gamma_0+Dq^2$ dependence. We find that $\Gamma_0$ is 
temperature-dependent. The relaxation time follows the Arrhenius law well. 
The presence of a relaxation mode associated with quasi-elastic scattering 
in PMN indicates that order-disorder behaviour plays an important r\^ole 
in the dynamics of diffuse phase transitions.
\end{abstract}


\maketitle

\noindent 
Enormous attention has been paid to relaxor 
ferroelectrics for more than 40 years due to their intriguing 
physical properties. A marking feature observed in relaxors 
is the frequency-dependent peak in the dielectric permeability 
$\varepsilon$ which typically extends 
over a few 100 degrees and does not 
link directly to the macroscopic changes of symmetry~\cite{cross}. 
Since many physical properties of relaxors exhibit anomalies in this 
temperature range, it was called "diffuse phase transition"~\cite{smol1}.
Despite intense investigations, however, the nature of the diffuse phase 
transition in these compounds remains 
unclear~\cite{kleemann, viehland, blinc}. 

The well-known crystal PbMg$_{1/3}$Nb$_{2/3}$O$_3$ (PMN) is a model relaxor. 
In PMN, the peak in the real part of the dielectric permeability  
$\varepsilon'$ appears around the mean Curie temperature 
T$_{cm}\sim$~270~K~\cite{smol1}. By application of an electric field 
it is possible to induce a structural phase transition at 
T$\sim210$~K~\cite{krainik}. Furthermore, at T$_d\sim 620$~K the 
temperature dependence of the optical refraction index deviates  
from the expected linear dependence~\cite{burns}. 
This result was explained by the appearance, 
far above T$_m$, of small polarized regions within the crystal~\cite{burns}. 
This polarization would exist on the scale of 
a few unit cells~\cite{burns} only, and hence called 'polar nanoregions'
(PNR). 
The concept of PNR became very popular to 
account for properties like the diffuse scattering observed by  neutron
scattering in relaxors of PbB$'_{1/3}$B$''_{2/3}$O$_3$ 
type~\cite{shirane1, shirane2, touluse1, seva1}. 

To obtain a theoretical description of a phase transition, 
an important point is to know whether this transition is primarily of  
\textit{displacement}, or of  \textit{order-disorder} type~\cite{cowley1}. 
Since PMN crystallizes in the perovskite structure where 
the B-site is disordered, it was expected that the diffuse phase 
transition is of displacement type. 
Within such a scenario a soft-phonon mode should be observed. 
Extensive search for the presence of a soft mode in PMN was performed both
by light 
(for a review see~\cite{siny1}) and neutron scattering. From light
scattering 
no soft mode has ever been found. On the other hand, neutron scattering 
experiments revealed a very complicated behaviour of the phonons 
in PMN~\cite{sb1, shirane3, sb2}. 
Firstly, a soft excitation was identified in PMN above T$_d$~\cite{sb1}, 
which is coupled with transverse acoustic phonon. 
Later, in Ref.~\cite{shirane3} a strong 
renormalization of the lowest transverse optic (TO) phonon branch was
observed starting from T$\sim$1100~K.    
Recently, a soft mode coupled to the 
lowest TO phonon was reported~\cite{sb2}. Obviously, the  
identification of a soft mode in PMN has not been definitely settled. 
In addition to the phonon modes a resolution-limited central peak (CP) is 
observed by neutron scattering in PMN without~\cite{sb1} and with an
applied electric field~\cite{sb3}. 
The timescale of CP is also at the
center of a controversy because it contradicts  
light scattering data~\cite{siny2}, where a broad quasielastic component was 
observed with a typical width of 2~meV. We note, that 
the search for a broad quasielastic component in 
related relaxor PbZn$_{1/3}$Nb$_{2/3}$O$_3$+8\%PbTiO$_3$ was not 
successful~\cite{kulda1} which again is in contradiction with  results of  
light scattering ({\it e.g.}~\cite{kojima1}). 
Hence, neutron scattering data speaks in favor of displacement
behaviour in relaxor ferroelectrics coexisting with 
a narrow resolution limited CP. On the contrary, light
scattering data suggest order-disorder behaviour in these compounds.

To solve these discrepancies 
between light and neutron scattering results and to show that 
order-disorder behaviour plays an important r\^ole in the lattice dynamics
of 
PMN we have studied the low-energy range of the vibration spectrum in this
crystal by inelastic cold-neutron scattering under improved resolution 
conditions with respect to the previous studies of refs.~\cite{sb1,sb2,sb3}. 
The measurements were carried out with the 
three-axis spectrometer TASP, located at the neutron spallation source SINQ 
at the Paul-Scherrer-Institute, Switzerland. A high-quality single 
crystal of PMN ($\sim$ 8~cm$^3$, mosaic $\sim 20'$) was mounted inside 
a closed-cycle refrigerator equipped with a small furnace. 
The crystal was aligned in the (h~h~l)-scattering plane.
The measurements were performed in the temperature range $100$~K - $450$~K.
The (002) reflection of pyrolytic 
graphite (PG) was used to monochromate and analyze the incident 
and scattered neutron beams, respectively. The spectrometer 
was operated in the constant final-energy mode with either  
$k_{f}=1.64$ \AA$^{-1}$ or $k_{f}=1.97$ \AA$^{-1}$ using a 
PG filter to remove higher-order wavelengths. 
The horizontal collimation was guide$-80'-80'-80'$. 
With that configuration the energy resolution as measured 
with a standard Vanadium sample is 0.4 meV for 
$k_{f}=1.97$ \AA$^{-1}$ and 0.2 meV for $k_{f}=1.64$ \AA$^{-1}$.
\begin{figure}[h]
  \includegraphics[width=0.4\textwidth, angle=0]{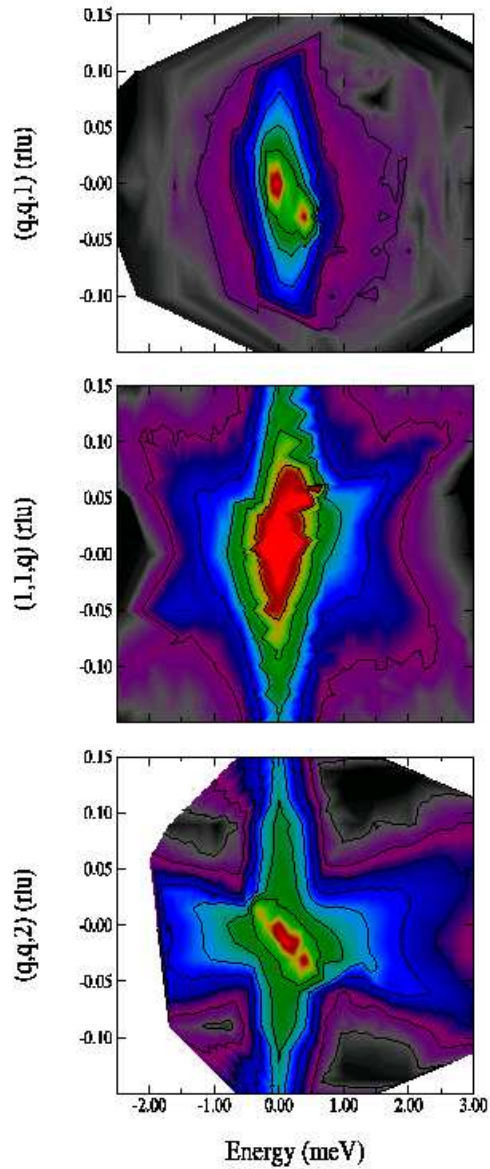}
  \caption{Color maps of neutron scattering intensity around the (001)
   ({\it top}), (110) ({\it middle}), and (002) ({\it bottom}) Bragg 
    positions at T = 300 K. Asymetry in the intensity contours 
    of neutron energy gain (negative energies) and neutron energy 
    loss (positive energies) parts of spectra is due to the 
    resolution effects. Note that the intensity is 
    given in a logarithmic scale. 1 rlu=1.555 $\AA^{-1}$.}
\label{maps}
\end{figure}
Figure~\ref{maps} shows ($q$, $\hbar\omega$) contour-maps of neutron
scattering spectra in the vicinity of Bragg positions (0~0~1), (1~1~0), 
and (0~0~2) at T~=~300~K, respectively. It is clear that all three 
figures differ from each other. For example, the intensity 
distribution in the (0~0~1) Brillouin zone (BZ) consists of a Bragg peak, 
a rather intense signal which is narrow in energy  
and broad in $q$, and a diffuse  signal which is 
broad both in $q$ and in $\hbar\omega$. In the (0,0,1) BZ no 
linear dispersion which could be associated 
with TA phonons is observed. On the other hand, one can easily see the
linear dispersion of TA phonons 
in the (0~0~2) BZ, in addition to the Bragg peak and to the intense signal 
narrow in energy. This signal corresponds to the central peak reported
before~\cite{sb1, sb2}. Qualitatively, the distribution of the intensity 
in the (1~1~0) BZ lies between what is observed in 
the (0~0~1) and (0~0~2) zones.
The typical peak intensity of a 
TA phonon measured close the (0~0~2) Brillouin zone center  
is $\sim$ 5000 counts at room temperature. 
It is well-known that the neutron intensity of an acoustic phonon scales
with the structure factor of the corresponding Bragg peak and with 
the square of the modulus of the scattering vector $Q^2$~\cite{lovesey1}. 
The structure factor of the (0~0~2) 
Bragg peak is 100 times larger then that of (0~0~1). Also, 
$Q^2$ increases by a factor of $\sim$~4 while measuring in the (0~0~2) BZ. 
Thus, the total decrease of intensity of the TA phonons in
(0~0~1) BZ with respect to (0~0~2) amounts to 400. Hence it
is impossible to detect the intensity of such a phonon with respect to the
diffuse
intensity which amounts to $\sim$200 counts in the (0~0~1) zone.
 
In this letter we will concentrate on the properties of the quasielastic
intensity
found in PMN and not on the behavior of the acoustic phonons as these have
been the subject of numerous studies~\cite{shirane1, sb1, shirane3, sb2,
shirane4}. 
In order to interpret the data quantitatively we decomposed the 
neutron scattering spectra $I(\mathbf{Q},\omega)$ in the following way:
\begin{equation}
\label{intensity}
I(\mathbf{Q},\omega)=[S_{CP}+S_{q-el.}+S_{DHO}] \otimes
R(\mathbf{Q},\omega)+B 
\end{equation}
\noindent where $S_{CP}$ refers to the resolution-limited central peak;
$S_{q-el.}$ is 
the quasi-elastic scattering shown in Fig.~1; $S_{DHO}$ describes the phonon
scattering. The symbol $\otimes$ stands for the convolution with the
spectrometer 
resolution function $R(\mathbf{Q},\omega)$~\cite{popa}. 
$B$ denotes the background level.

We found that the data could be modeled using  
 \begin{equation}
 \label{CP}
 S_{CP}=A(q)\delta(\omega),
 \end{equation}
 \begin{equation}
 \label{quasiel}
 S_{q-el}=I(q)\frac{\omega}{1-\exp(-\omega/T)}
 \frac{1}{\pi}\frac{\Gamma_q}{\omega^2+\Gamma_q^2},
 \end{equation}
and
 \begin{equation}
 \label{dho}
  S_{DHO}=W(q)\frac{\omega}{1-\exp(-\omega/T)}
  \frac{\gamma_{q}}
  {(\omega^2-\Omega_{q}^2)^2+\omega^2\gamma_{q}^2}.
 \end{equation}

\begin{figure}[h]
  \includegraphics[width=0.475\textwidth]{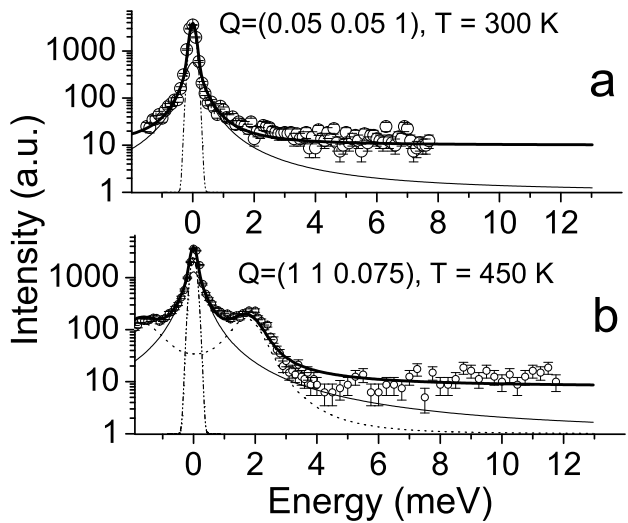}
  \caption{Typical examples of constant-Q scans in different BZ.
           The bold line shows the results of data analysis 
           as described in the text. The solid line shows the 
           contribution of the quasielastic component, the 
           dash-dotted line the narrow central peak and the dotted 
           line the TA phonon.}
\label{scans}
\end{figure}

\begin{figure}[h]
  \includegraphics[width=0.30\textwidth]{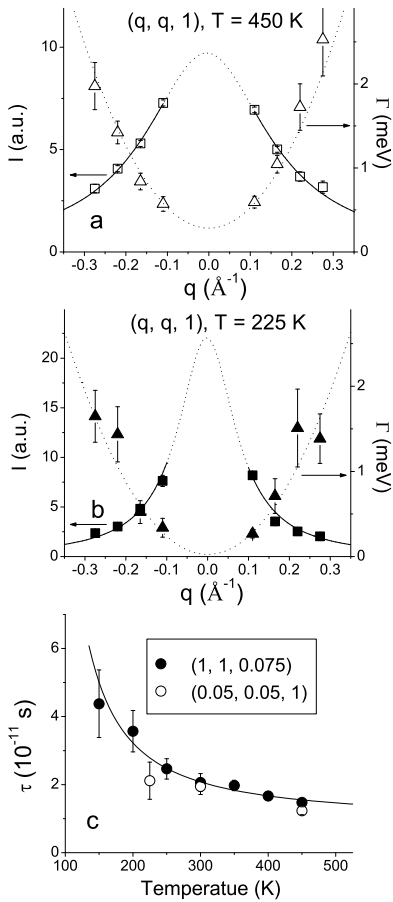}
  \caption{{\bf a} and {\bf b} q- dependence of integrated intensity
           (squares) and of damping (triangles) of the QE component at 
           $T$~=~450~K and 225~K, respectively .
           {\bf c} Temperature dependence of the relaxation
           time $\tau$ associated with the quasielastic component.
           The solid line is a fit to the data with the Arrhenius law.
           }
\label{dep}
\end{figure}

\noindent Here $\delta(\omega)$ is the Dirac delta function; $\Gamma_q$ and
$\gamma_q$ are the half-width at half maximum (HWHM) of the Lorentzian and
damped-harmonic oscillator function, respectively.
$A(q)$, $I(q)$ are
the q-dependent susceptibilities of the corresponding excitations and
$W(q)$ is the TA dynamical structure factor.
The renormalized dispersion of the TA phonons is given by 
$\Omega_{q}=\sqrt{\omega_{q}^2+\Gamma_{q}^2}$ where $\omega{_q}$ is the
frequency of a damped oscillation. In our description, $\Omega_{q}$ 
is the physically relevant quantity~\cite{bruno}. Since the 
measurements are restricted to small values of $q$,
we used a linear dispersion for the TA phonon.
The validity of such an approximation is clear from 
the intensity distribution shown in Fig.~1. 
Fig.~\ref{scans} shows typical constant-Q scans in
the (0~0~1) and (1~1~0) BZ, respectively. 
Comparison between Figs.~\ref{scans}a and \ref{scans}b shows that phonons 
are visible 
at $\hbar\omega=\pm~1.6$~meV for Q=(1~1~0.075). On the contrary there is 
no phonon scattering at Q=(0.05~0.05~1).  
Hence, to analyze the data in the (0~0~1) BZ, we fixed $W(q)$ to zero.  
The solid lines in Fig.~\ref{scans} show that the data are well represented
by
equation~\ref{intensity}. 
\noindent Figs.~\ref{dep}a,b show the q- dependence of the integrated
intensity and of the damping of the quasielastic scattering observed in the 
(0~0~1) BZ at $T$ = 450~K and 225~K.
The quasielastic scattering can be approximated by a Lorentzian line-shape
in q-space. Then the correlation length is $\xi$~=~6~\AA~at T~=~450~K, 
and $\xi$~=~12~\AA~at T~=~225~K ($\xi$=1/HWHM of the Lorentzian). The 
damping of the QE component is found to behave like 
$\Gamma (q)=\Gamma_0+Dq^2$. At $T$~=~450~K the parameters are 
$\Gamma_0=0.28\pm0.01$~meV and $D=24.5\pm7.3$~meV$\cdot$\AA$^{2}$ and 
at $T$~=~225~K we obtain $\Gamma_0=0.02\pm0.01$~meV and 
$D=21.3\pm8.9$~meV$\cdot$\AA$^2$.  
The absence at all $q$ of an intrinsic energy width of CP implies that the 
associated fluctuations are very slow.  
Further, the intensity of the QE scattering does not scale 
with the narrow CP. In the (0~0~2) BZ (Fig.\ref{maps} {\it bottom}) 
the CP is very intense, but we were unable to detect QE component 
in this BZ. On the contrary, in the (0~0~1) BZ the narrow CP is
significantly weaker and the QE component is clearly seen. 
The different q-dependences of the energy widths and behavior of the
relative intensities of the CP and  quasi-elastic scattering indicate 
that these processes are of different origins.  

We now turn to the temperature behaviour of the relaxation time  
$\tau={1/\Gamma}$ of the QE-mode measured in the (0,0,1) and (1,1,0) BZ,
respectively, and shown in Fig.~\ref{dep}c. It is obvious, that the
temperature dependence of $\tau$ is very similar in both zones. We 
found that $\tau$ is well described by the Arrhenius law 
$\tau (T)=\tau_0\cdot \exp{[E_a/k_B\cdot T]}$. The solid line
in Fig.~\ref{dep}b corresponds to the parameters 
$\tau_0=8.6\pm0.1\cdot10^{-12}$~(s) and $E_a=0.023\pm0.0002$~(eV). 
While these parameters are reasonable, they have to be taken with care 
as quantities like the refraction index~\cite{burns},
the specific heat~\cite{moriya}, and the diffuse 
scattering~\cite{shirane1, sb4} exhibit anomalous behaviours in this 
temperature range. This implies that the structure
of PMN undergoes significant changes on a local scale, which in turn 
should affect the value of $E_a$. Despite this uncertainty, the fact remains 
that $\tau(T)$  is connected to a thermally activated behaviour, which
allows us to speculate about the origin of the relaxation mode.
It is well-known, that in relaxors, Pb ions are shifted 
from their ideal positions in the perovskite structure~\cite{sb4}. 
Recently it was shown that the probability density function $\rho(R)$ 
for Pb in PMN has more than a single maximum below T$_d$~\cite{sb5}. 
Therefore, it is reasonable to assume that at high temperature the Pb ions  
move between maxima of $\rho(R)$. 
In that case when the temperature is low enough with respect with
the potential barrier, the hopping frequency $1/\tau$ becomes small.
This gives rise to a narrowing in the energy-width of the QE component,
which, within the limitations of the present experiment, 
is resolution limited at $T=100$~K in PMN. 

There is a number of models which describe the presence of 
a central peak in the vicinity of a structural phase transition (for a 
review of earlier papers see~\cite{cowley2}). However, PMN does not 
undergo any structural phase transition without application of an electric 
field. To link the coexistence of a central peak with quasielastic 
scattering in a broad temperature range might be of importance to 
understand the diffuse phase transition in relaxor ferroelectrics.

To summarize, we have carried out high-resolution neutron scattering
measurements in the relaxor ferroelectric PMN. The low-energy spectra 
exhibit two components associated with two different timescales. 
One is the resolution-limited central
peak. The other component is quasi-elastic and appears to be related to   
ionic motion of thermally activated character.
The damping of this relaxation mode follows a $q^2$ dependence. 
The presence of a relaxation mode associated 
with quasi-elastic scattering in PMN indicates that order-disorder
behaviour plays an important r\^ole in the dynamics of diffuse phase
transitions. It will be interesting to see if this relaxation mode, which
was also observed in PMT~\cite{seva1}, is a common feature of all relaxor 
ferroelectrics.

\begin{acknowledgments}
The authors would like to thank Prof. R.A. Cowley for invaluable discussions. 
This work was performed at the spallation neutron source SINQ, 
Paul Scherrer Institut, Villigen (Switzerland) and was partially 
supported by RFBR grant 02-02-17678. 
\end{acknowledgments}

%


\begin{references}
\bibitem{cross} L.E. Cross, Ferroelectrics, \textbf{76}, 241 (1987).
\bibitem{smol1} G.A. Smolenskii {\it et al.}, {\it Ferroelectrics and
Related Materials}, (Gordon and Breach, NY, 1984).
\bibitem{kleemann} V. Westphal, W. Kleemann, and M. D. Glinchuk,
Phys. Rev. Lett. \textbf{68}, 847 (1992).
\bibitem{viehland} D. Viehland {\it et al.},
Phys. Rev. B \textbf{46}, 8003 (1992).
\bibitem{blinc} R. Blinc {\it et al.},
Phys. Rev. Lett. \textbf{83}, 424 (1999).
\bibitem{krainik}G. Schmidt {\it et al.}, Krist. und Tech. \textbf{15},
1415 (1980).
\bibitem{burns} G. Burns and B.A. Scott, Solid State Commun. \textbf{13},
423 (1973).
\bibitem{shirane1} K.Hirota {\it et al.}, Phys. Rev. B \textbf{65}, 
104105 (2002).
\bibitem{shirane2} S. Wakimoto {\it et al.}, Phys. Rev. B \textbf{66},
224102 (2002).
\bibitem{touluse1} D. La-Orauttapong {\it et al.}, Phys. Rev. B \textbf{64},
212201 (2001).
\bibitem{seva1} S.N. Gvasaliya, B. Roessli, and S.G Lushnikov, Europhys.
Lett. \textbf{63}, 303 (2003).
\bibitem{cowley1} R.A. Cowley, Adv. in Phys. \textbf{29}, 1 (1980).
\bibitem{siny1} I.G. Siny {\it et al.}, Ferroelectrics
\textbf{226}, 191 (1997).
\bibitem{sb1} A. Naberezhnov {\it et al.}, Eur. Phys. J. B
\textbf{11}, 13 (1999).
\bibitem{shirane3} P. M. Gehring {\it et al.},
Phys. Rev. Lett. \textbf{87}, 277601 (2001).
\bibitem{sb2} S. B. Vakhrushev and S. M. Shapiro,
Phys. Rev. B \textbf{66}, 214101 (2002).
\bibitem{sb3} Yu.O. Chetverikov {\it et al.},
Appl. Phys. A \textbf{74}, S989 (2002).
\bibitem{siny2}I.G Siny {\it et al.},
Phys. Rev. B \textbf{56}, 7962 (1997).
\bibitem{kulda1} J. Hlinka {\it et al.}, J. Phys.: Condens. 
Matter \textbf{15}, 4249 (2003).
\bibitem{kojima1} F. Jiang and S. Kojima,
Ferroelectrics \textbf{266}, 19 (2002).
\bibitem{lovesey1} {\it e.g.} S.W. Lovesey, {\it Thermal neutron
scattering},
(Oxford University Press, UK, 1984).
\bibitem{shirane4} S. Wakimoto {\it et al.}, Phys. Rev. B 
\textbf{65}, 172105 (2002).
\bibitem{popa} M. Popovichi,
Acta Cryst. A \textbf{31}, 507 (1975).
\bibitem{bruno} B. F{\aa}k and B. Dorner, Physica B \textbf{234}, 1107 (1997).
\bibitem{moriya} Y. Moriya {\it et al.}, Phys. Rev. Lett
\textbf{90}, 205901 (2003).
\bibitem{sb4} B. Dkhil {\it et al.}, 
Phys. Rev. B \textbf{65}, 024104 (2002). 
\bibitem{sb5} S. B. Vakhrushev and N.M. Okuneva, AIP 
Conference Proceedings \textbf{626}, 117 (2002).
\bibitem{cowley2} A.D. Bruce and R.A. Cowley, 
Adv. in Phys., \textbf{29}, 219 (1980).
\end{references}
\end{document}